\documentclass[12pt]{article}
\usepackage[totalwidth=460truept,totalheight=605truept]{geometry}
\usepackage{latexsym,graphicx,amsfonts,amssymb,amsmath,xcolor}
\usepackage[hypertexnames=false,hidelinks]{hyperref}

\global\arraycolsep=1truept

\begin{document}

\begin{center}
{\huge \textbf{Purely Virtual Particles}}

\vskip.8truecm

{\huge \textbf{versus Lee-Wick Ghosts:}}

\vskip1truecm

{\huge \textbf{Physical Pauli-Villars Fields,}}

\vskip.8truecm

{\huge \textbf{Finite QED,\ and Quantum Gravity}}

\vskip.8truecm

\textsl{Damiano Anselmi}

\vskip.1truecm

{\small \textit{National Institute of Chemical Physics and Biophysics, R\"{a}%
vala 10, Tallinn 10143, Estonia}}

{\small \textit{Dipartimento di Fisica \textquotedblleft
E.Fermi\textquotedblright , Universit\`{a} di Pisa, Largo B.Pontecorvo 3,
56127 Pisa, Italy}}

{\small \textit{INFN, Sezione di Pisa, Largo B. Pontecorvo 3, 56127 Pisa,
Italy}}

{\small damiano.anselmi@unipi.it}

\vskip.5truecm

\textbf{Abstract}
\end{center}

We reconsider the Lee-Wick (LW) models and compare their properties to the
properties of the models that contain purely virtual particles. We argue
against the LW premise that unstable particles can be removed from the sets
of incoming and outgoing states in scattering processes. The removal leads
to a non-Hermitian classical limit, besides clashing with the observation of
the muon. If, on the other hand, all the states\ are included, the LW models
have a Hamiltonian unbounded from below or negative norms. Purely virtual
particles, on the contrary, lead to a Hermitian classical limit and are
absent from the sets of incoming and outgoing states without implications on
the observation of long-lived unstable particles. We give a \textit{vademecum%
} to summarize the properties of most options to treat abnormal particles.
We study a method to remove the LW\ ghosts only partially, by saving the
physical particles they contain. Specifically, we replace a LW\ ghost with a
certain superposition of a purely virtual particle and an ordinary particle,
and drop only the former from the sets of the external states. The trick can
be used to make the Pauli-Villars fields consistent and observable, without
sending their masses to infinity, or to build a finite QED, by tweaking the
original Lee-Wick construction. However, it has issues with general
covariance, so it cannot be applied as is to quantum gravity, where a
manifestly covariant decomposition requires the introduction of a massive
spin-2 multiplet.

\vfill\eject

\section{Introduction}

\label{intro}\setcounter{equation}{0}

In the late 1960s Lee and Wick proposed a way to give sense to models that
contain fields with negative kinetic terms \cite{leewick,LWQED}. A key point
of their idea is that \textquotedblleft abnormal\textquotedblright\
particles do not belong to the spectrum of asymptotic states, as long as
they are unstable. In their approach, it is sufficient that all the stable
particle states have positive square lengths. The purpose of Lee and Wick
was to provide a unitary $S$ matrix in the subspace of stable states, by
extending the previous results on unitarity \cite%
{cutkosky,Veltmanstable,thooft,diagrammar,diagrammatica}. In this paper, we
reconsider the Lee-Wick (LW) models, concentrating on the treatment of
unstable particles.

The muons are unstable elementary particles that can be observed directly
before they decay. Tauon traces can also be observed in special situations.
Composite long-lived particles are more common, but their relatively long
lifetimes can be due to their compositeness. The other elementary particles
are stable, short-lived, or confined. Moreover, the resonances can in
principle be boosted enough and detected as particles before they decay. In
light of these remarks, it does not seem so justified to remove a particle
from the set of asymptotic states just because it is unstable.

In the context of the Lee-Wick models, it is actually sufficient to remove
the abnormal particles from the sets of the external states and keep the
physical particles as usual. The advantage of this modified removal option
is that it does not clash with the observation of the muon. Nevertheless, it
leads to an unacceptable classical limit.

The classical limit is given by the tree diagrams that have physical
particles in the external legs and no physical particles in the internal
legs. We can also define a \textquotedblleft reduced\textquotedblright\
action, which is the effective action\ obtained by integrating out the
abnormal particles. It collects all the diagrams that have physical
particles in the external legs and no physical particles in the internal
legs, and includes the loops of abnormal particles as effective vertices. If
we want to define a fundamental theory by removing particles, the reduced
action should be seen as the \textquotedblleft classical\textquotedblright\
action of that fundamental theory.

The abnormal particles propagating in the internal legs generate nonlocal,
acausal, non-Hermitian effective self-interactions among the physical
particles. Violations of locality and causality in the classical action are
not excluded by the requirements of internal consistency, as long as they
are microscopic and compatible with data, which occurs (for example) if the
masses of the abnormal particles are sufficiently large. On the other hand,
a classical Lagrangian with a nonvanishing imaginary part is troubling. The
simplest explanation for such an instance is that something has been
provisionally integrated out, which is precisely what is going on in the
case we are considering. If we reinstate the missing entity (which is the
Lee-Wick abnormal particle) as an independent degree of freedom, we remove
the imaginary part of the classical Lagrangian, but go back to the initial
problem of negative kinetic terms (free Hamiltonian unbounded from below).
This either-or situation is the trouble with the Lee-Wick models.

For most purposes, the muon can be treated as a stable particle, since its
width is very small (around $10^{-19}$GeV). If we resum the muon
self-energies into the dressed propagator, as is commonly done for
resonances, we find that the theory predicts no muon observation \cite%
{FselfE}. The reason is that we cannot observe an unstable particle with
infinite resolving power on the energy: such an instance would violate the
energy-time uncertainty principle. Once the energy resolution $\Delta E$ of
the experimental setup is inserted explicitly, the problem disappears \cite%
{FselfE}. We can argue in a similar way for every resonance, if we imagine
to boost it enough to make it detectable as a particle. Indeed, the
resonances and the muon just differ by the magnitudes of their widths. We
conclude that in a sound theoretical framework unstable particles should be
included among the external states.

\bigskip

In this paper, we compare the Lee-Wick idea to several other options,
including physical particles, ordinary ghosts and purely virtual particles.
We also consider the effects of the removal of those from the sets of
incoming and outgoing states. By \textquotedblleft ghost\textquotedblright\
we mean a degree of freedom that appears with a negative kinetic term in the
classical Lagrangian.

Purely virtual particles, or fake particles, or \textquotedblleft
fakeons\textquotedblright , are based on a new diagrammatics \cite%
{diagrammarMio}. They allow us to formulate a consistent theory of quantum
gravity \cite{LWgrav}, which is experimentally testable due to its sharp
prediction of the tensor-to-scalar ratio in inflationary cosmology \cite{ABP}%
. They can also be used to search for new physics beyond the standard model,
by evading common constraints in collider phenomenology \cite{Tallinn1} and
offering possible resolutions of discrepancies with data \cite{Tallinn2}.
The only requirement is that fakeons are massive and nontachyonic. Their
diagrammatics can be implemented in softwares like FeynCalc, FormCalc,
LoopTools and Package-X \cite{calc}.

Unlike the LW abnormal particles (which we call \textquotedblleft LW
ghosts\textquotedblright\ from now on), fakeons lead to a Hermitian
classical limit and a Hermitian reduced action. Their absence from the sets
of incoming and outgoing states has no implication on the observation of
long-lived unstable particles. The reason is that the fakeons are purely
virtual. Instead, the LW ghosts are not purely virtual, which is why they
leave an imaginary remnant in the classical limit, once they are removed.

After reconsidering the Lee-Wick construction, we formulate a procedure that
is as close as possible to the idea of removing the LW ghost from the sets
of external states only partially and save the physical degree of freedom it
contains. Specifically, we switch to a theory of particles and fakeons by
replacing the LW ghost with a certain superposition of a fakeon and an
observable particle, and remove only the former. The trick works with
neutral matter fields and can be used to make the Pauli-Villars fields
consistent, and observable, without sending their masses to infinity. It
also allows us to build a finite QED, by overcoming the difficulties of the
original Lee-Wick construction. In quantum gravity, the method could lead to
an extra (observable) massive spin-2 particle. However, a number of
unresolved issues with general covariance (and gauge invariance) show that
it cannot be applied to gravity as is.\ A covariant decomposition can be
achieved by adding a massive spin-2 multiplet (which can be done in a
unitary and renormalizable way as explained in \cite{HS}). However, this
procedure just gives the theory of \cite{LWgrav} coupled to matter in a
peculiar way.

We do not cover all the proposals available in the literature about ghosts.
Among the missing ones, we mention the PT (parity and time reversal)
symmetric approach of Berends and Manheim \cite{berends}.

\bigskip

The removal of degrees of freedom from the incoming and outgoing states is
consistent only if it is compatible with unitarity, in which case we call it
\textquotedblleft projection\textquotedblright\ and call the action
\textquotedblleft projected action\textquotedblright . The fakeon projection
is compatible with unitarity order by order (and diagram by diagram) in the
perturbative expansion (see for example \cite{diagrammarMio}). The removal
of unstable particles (which we call Veltman's projection, see below) is
compatible with unitarity in a semi-perturbative approach, because the
self-energies of unstable particles must be resummed into their dressed
propagators. After this diagrammatic reorganization, it is also valid
diagram by diagram.

Since the fakeon approach is perturbative, we require the Hamiltonian to be
bounded from below in the free-field limit (in flat space), both classically
and at the quantum level. Once a particle is projected away, it is no longer
relevant to the issue, because it disappears from the free-field limit. We
have no way to say whether the Hamiltonian is bounded or not in the complete
theory. In simple models, the nonlocalities surviving the classical limit
are diluted by the fakeon projection into an asymptotic series of
perturbative corrections \cite{FLRW}. In other cases, they affect only high
orders, where they compete with the quantum corrections, which are nonlocal
anyway. For example, in \cite{FakeOnScalar} it is shown that, in primordial
cosmology, the fakeon projection leaves the theory practically local for
various orders of the perturbative expansion.

The paper is organized as follows. In section \ref{preliminaries} we discuss
Veltman's projection and the issue of unitarity with unstable particles. In
section \ref{basics} we compare various options for the quantization of
fields with negative kinetic terms. In section \ref{loops} we briefly recall
how such fields are treated inside the loop diagrams. In section \ref{PV} we
present the trick that makes the Pauli-Villars physical by means of fakeons.
In section \ref{QED} we apply it to build a finite QED. In section \ref{QG}
we discuss the obstacles we meet when we apply the same method to quantum
gravity. Section \ref{conclusions} contains the conclusions.

\section{Veltman's projection}

\label{preliminaries}\setcounter{equation}{0}

A result due to Veltman states that the $S$ matrix constructed with the
dressed propagators and connecting stable particle states only is unitary 
\cite{Veltmanstable}. Because of this, unstable particles can be
consistently dropped from the sets of incoming\ and outgoing\ states of the
scattering processes. We call this removal Veltman's projection. The $S$
matrix obtained from it is called reduced (or projected) $S$ matrix\ and
denoted by $S_{r}$.

Veltman's result $S_{r}^{\dagger }S_{r}=1$ follows from the common proofs of
perturbative unitarity by means of cut diagrams \cite%
{Veltmanstable,cutkosky,thooft,diagrammar,diagrammatica}. When an unstable
particle is projected away by means of Veltman's projection, it generates
effective nonlocal, non-Hermitian interactions among the other particles
(see, for example, formula (\ref{none}) below). In general, non-Hermitian
interactions are problematic for unitarity, but in the case of Veltman's
projection they are precisely what makes the unitarity equation $%
S_{r}^{\dagger }S_{r}=1$ hold true. At the end of this section, we briefly
recall how this happens and also show that Veltman's projection preserves
CPT\ invariance.

\bigskip

Veltman considered stable and unstable particles. To apply Veltman's
projection to ghosts, we should first ensure that Veltman's results extend
to them. There are various options to treat ghosts at the quantum level.

The simplest possibility, which we call $i\epsilon $ ghost, is the standard
quantization by means of the Feynman $i\epsilon $ prescription, which means
that we choose the free propagator%
\begin{equation}
-\frac{i}{p^{2}-M^{2}+i\epsilon }  \label{iep}
\end{equation}%
and integrate on Minkowski spacetime as usual. In ref. \cite{FselfE} it has
been shown that the dressed propagators of the $i\epsilon $ ghosts do not
make sense close to the peaks, because the resummation of the perturbative
expansion does not exist there. Having no knowledge about the
nonperturbative sector of the theory, Veltman's projection cannot be applied
to the $i\epsilon $ ghosts. This is not a big deal, since the $i\epsilon $
ghosts violate unitarity.

Other options to quantize ghosts have different properties. The second
possibility, which we call $-i\epsilon $ ghost, is to choose the free
propagator%
\begin{equation}
-\frac{i}{p^{2}-M^{2}-i\epsilon }.  \label{-iep}
\end{equation}%
Then we cannot integrate on Minkowski spacetime, because if we do so we run
into the consistency problems described in ref. \cite{wheelerons}, which
means nonlocal divergent parts, exchanges of roles between the usual
thresholds and the pseudothresholds, instabilities, violations of unitarity,
etc. Lee and Wick proposed a different set of rules for handling (\ref{-iep}%
) in Feynman diagrams, which must be combined with the Cutkosky \textit{et
al.} (CLOP) prescription \cite{CLOP} and possibly other rules, to solve the
ambiguities mentioned in \cite{CLOP} (see also \cite{LWformulation}). For
the purposes of this paper, we can just assume that a complete set of rules
does exist. At the end, the $-i\epsilon $ ghost turns into a new type of
object, which we call a LW ghost.

More importantly, the idea of Lee and Wick is to arrange the model so that
the interactions make the ghost unstable, to apply Veltman's projection to
it. To this purpose, we note that the dressed propagator makes sense, even
close to the peak, where it reads\footnote{%
Formula (\ref{dre2}) is correct as is for legs that disconnect the diagram
once they are broken. Inside loops we must use the rules mentioned
previously.} 
\begin{equation}
-\frac{iZ}{p^{2}-M_{\text{gh}}^{2}-i(\epsilon +M_{\text{gh}}\Gamma )}.
\label{dre2}
\end{equation}%
Here, $\Gamma $ is a positive width, $M_{\text{gh}}^{2}=M^{2}+\Delta M^{2}$
is the \textquotedblleft physical\textquotedblright\ mass squared and $Z$ is
the normalization factor. The $\epsilon $ prescription is there to show that
the corrections proportional to $\Gamma $ have the same sign. According to
the arguments of \cite{FselfE}, it is correct to extend the resummation of
the self-energies from the convergence region to the peak region by means of
analyticity.

To summarize, Lee and Wick get rid of ghosts by turning them into LW ghosts
and arranging the model so that they are unstable or become so dynamically,
to build a unitary reduced $S$ matrix $S_{r}$ \`{a} la Veltman.

Although Veltman's result is correct, it does not suggest that we \textit{%
should} drop unstable particles from the physical spectrum\footnote{%
Actually, Veltman seems to suggest precisely that in \cite{Veltmanstable},
by saying that it is an undesirable feature of perturbation theory to have
unstable particles among the asymptotic states. Our position, instead, is
that a theory of scattering where processes end at the end of time is not
satisfactory.}. It simply proves that if we drop them, we get a unitary
reduced scattering matrix $S_{r}$. The problem with $S_{r}$ is that it turns
a blind eye to the experimental observation of the muon.

Normally, the incoming\ and outgoing\ states\ of a scattering process are
assumed to be at $t=-\infty $ and $t=+\infty $, respectively. This is a
nonrealistic simplification, useful to derive general formulas. A more
realistic assumption is $\bar{\Delta}t\ll \Delta t<\infty $, where $\Delta t$
denotes the time separation between the incoming\ and outgoing\ states and $%
\bar{\Delta}t$ is the duration of the interactions. This leaves room for
including long-lived unstable particles, by assuming that $\Delta t$ is
smaller than the lifetimes of some of them. Once the scattering process ends
and the outgoing\ particles fly away, there is no reason why we should wait
till they decay, if we can catch them on the fly.

To establish an unambiguous terminology, we talk about \textquotedblleft
physical\textquotedblright\ spectra when we include everything we can
physically observe, in practice or in principle. Clearly, the muon is
included, among the other unstable particles. With the word
\textquotedblleft asymptotic\textquotedblright , we mean the same, i.e., $%
\Delta t\gg \bar{\Delta}t$. Thus, the muon is also included in the set of
\textquotedblleft asymptotic states\textquotedblright\ and is part of the
asymptotic spectra. Incoming\ and outgoing\ states that are literally taken
at $t=-\infty $ and $t=+\infty $ will be called \textquotedblleft strictly
asymptotic states\textquotedblright .

Given that we never see resonances like the $Z$ boson, one could ask why we
should include them in the physical spectra. The reason is that a
fundamental theory should be able to cover all the situations, including the
ones that are currently out of reach experimentally.

Although Veltman's projection is not acceptable for physical particles,
because it forces us to drop the muon from the physical spectrum of the
standard model, we could accept a restricted form of it, by applying it to
the LW ghosts only. The restricted option is compatible with the
observations of long-lived unstable particles. Nevertheless, Veltman's
projection has another problem, which concerns the classical limit.

Every unstable particle becomes stable in the classical limit, by
definition. If we ignore unstable particles as asymptotic states at the
quantum level, the classical limit cannot resuscitate them. This means that
the reduced $S$ matrix $S_{r}$ does not correspond to an acceptable
classical Lagrangian, typically because the latter turns out to be
non-Hermitian.

Thus, even if there existed no muon in nature, or we applied Veltman's
projection to the LW ghosts only, the model would still not be good enough
to define a fundamental quantum field theory, although it could be
acceptable in the realm of effective field theories.

The theories with fakeons do not have these problems, because they are
defined in a radically different way. In particular, fakeons are purely
virtual particles, so they do not need to be unstable and decay\ to be
removed from the physical spectrum, which they never enter\ in the first
place. There is no implication on the observation of unstable long-lived
particles like the muon. The classical limit is described by a Hermitian
Lagrangian, which collects, after the projection, anti-Hermitian effective
vertices. Moreover, the diagrammatic analysis of \cite{diagrammarMio} shows
that all the effective vertices given by the 1PI\ diagrams with no fakeons
in the external legs and no physical particles in the internal legs are
anti-Hermitian, even if they close loops. Thus, the reduced action is
Hermitian and the CPT theorem holds after the projection.

The reason why the LW ghosts leave an imaginary remnant in the classical
limit is that they are not purely virtual. What Lee and Wick suggest, i.e.,
assume that they are unstable and drop them from the physical spectrum, does
not remove them completely.

\subsection{Unitarity, Hermiticity and CPT\ invariance}

Normally, the unitarity equation $S^{\dagger }S=1$ is proved by means of
\textquotedblleft cutting equations\textquotedblright , which are identities%
\begin{equation}
G+\bar{G}+\sum_{c}G_{c}=0,  \label{cuto}
\end{equation}%
among cut and uncut diagrams. Specifically, one rewrites $S^{\dagger }S=1$
as the optical theorem $iT-iT^{\dagger }+T^{\dagger }T=0$, where $S=1+iT$.
Then, $G$ is the uncut diagram and stands for $iT$, $\bar{G}$ is its complex
conjugate and stands for $-iT^{\dagger }$, while $G_{c}$ are the cut
diagrams, which are obtained by cutting internal lines, and stand for $%
T^{\dagger }T$. The cut propagators encode the on-shell content of the full
propagators.

Let us see how the unitarity equation $S_{r}^{\dagger }S_{r}=1$ works after
Veltman's projection. We can build $S_{r}$ in two ways. The straightforward
method is to quantize the classical, unprojected Lagrangian $\mathcal{L}$ as
usual, build the (unprojected) $S$ matrix from it and perform Veltman's
projection at the very end. The second method is to work out the projected
Lagrangian $\mathcal{L}^{\text{V}}$ right away and then derive $S_{r}$ from $%
\mathcal{L}^{\text{V}}$. Then, however, $\mathcal{L}^{\text{V}}$ contains
effective non-Hermitian interactions due to the removal of the unstable
particles. We show that these effective interactions make unitarity work as
desired.

The projection $\mathcal{L}\rightarrow \mathcal{L}^{\text{V}}$ is obtained
in two steps. First, one builds the effective vertices, which are given by
the one-particle irreducible (1PI) diagrams (generated by $\mathcal{L}$)
that have stable particles in the external legs and no stable particles in
the internal legs. Second, the self-energies of the unstable particles are
resummed into their dressed propagators. In the end, $\mathcal{L}^{\text{V}}$
is made of \textquotedblleft dressed effective vertices\textquotedblright .
In some sense, it is semi-nonperturbative. The point is that its vertices
are not Hermitian, in general. So, how can the $S$ matrix $S_{r}$ obtained
from $\mathcal{L}^{\text{V}}$ be unitary?

In the algebraic approach of ref. \cite{diagrammarMio}, it is simple to
prove that a non-Hermitian classical Lagrangian $\mathcal{L}^{\text{V}}$
leads to a generalized version of the cutting equations of the form%
\begin{equation}
G+\bar{G}+\sum_{c}G_{c}+\sum_{c^{\prime }}G_{c^{\prime }}=0,  \label{cut}
\end{equation}%
where $G$ and $\bar{G}$ are as above, $G_{c}$ are the cut diagrams obtained
by cutting internal (stable-particle) lines and $G_{c^{\prime }}$ are
additional cut diagrams, obtained by cutting the non-Hermitian vertices as
well. Normally, the extra terms $G_{c^{\prime }}$ quantify the violations of
unitarity, because they have no interpretation in the unitary equation $%
S^{\dagger }S=1$. However, in the case of Veltman's projection, they are
precisely what is needed to interpret the identities (\ref{cut}) as the
correct diagrammatic versions of $S_{r}^{\dagger }S_{r}=1$. The reason is
that the cut vertices of $G_{c^{\prime }}$ describe the decays of the
unstable particles that have been projected away, which are not included in
the diagrams $G_{c}$. More details and the diagrammatic analysis of the
extra terms can be found in Veltman's paper \cite{Veltmanstable}.

Another issue is the CPT\ theorem after Veltman's projection. If the
unprojected theory is CPT\ invariant, the projected theory described by $%
S_{r}$ should be CPT invariant as well. The trouble is, again, that $%
\mathcal{L}^{\text{V}}$ is not Hermitian. If we want to have CPT\ invariance
after the projection, we must treat the effective vertices of $\mathcal{L}^{%
\text{V}}$ in a particular way under that symmetry.

Specifically, let $S(\mathcal{L},\epsilon )$ denote the $S$ matrix built
from a local Lagrangian $\mathcal{L}$ with Feynman's $i\epsilon $
prescription. Then, $S(-\mathcal{L}^{\dagger },-\epsilon )$ is the conjugate
matrix $S^{\dagger }$. CPT\ invariance is the statement that $S(\mathcal{L}%
^{\dagger },\epsilon )=S$, or $S(-\mathcal{L},-\epsilon )=S^{\dagger }$,
which is true if $\mathcal{L=L}^{\dagger }$. If we take $\mathcal{L}^{\text{V%
}}$ as the Lagrangian, we have $S(\mathcal{L}^{\text{V}},\epsilon )=S_{r}$,
where $\epsilon $ refers to the stable particles only. The point is that $%
\mathcal{L}^{\text{V}}$ also depends on the Feynman prescription (for the
unstable particles projected away). So, $S(-\mathcal{L}^{\text{V}}(\epsilon
),-\epsilon )\neq S_{r}^{\dagger }$. Nevertheless, we have the identities $(%
\mathcal{L}^{\text{V}}(\epsilon ))^{\dagger }=\mathcal{L}^{\text{V}%
}(-\epsilon )$ and $S(-\mathcal{L}^{\text{V}}(-\epsilon ),-\epsilon
)=S_{r}^{\dagger }$, which can be interpreted as the CPT\ theorem for $S_{r}$%
.

\section{Basic quantization options}

\label{basics}\setcounter{equation}{0}

In this section we compare the quantizations of physical particles, ghosts
and purely virtual particles and emphasize their basic properties, also in
relation with Veltman's projection, when it applies. We concentrate on the
tree diagrams, the classical limit and the dressed propagators. In the next
section we consider the loop diagrams.

We start from the Lagrangian%
\begin{equation}
\mathcal{L}_{\text{cl}}=\frac{1}{2}(\partial _{\mu }\varphi )(\partial ^{\mu
}\varphi )-\frac{m^{2}}{2}\varphi ^{2}+\mathcal{L}_{\phi }-g\varphi ^{2}\phi
-\Lambda \phi  \label{Lcl}
\end{equation}%
in four spacetime dimensions, which couples a physical particle $\varphi $
to some other type of particle $\phi $, to be defined below, with free
Lagrangian%
\begin{equation}
\mathcal{L}_{\phi }=\frac{\rho }{2}\left[ (\partial _{\mu }\phi )(\partial
^{\mu }\phi )-M^{2}\phi ^{2}\right] .  \label{L}
\end{equation}%
For the time being, we assume $M>2m$ and $\rho =\pm 1$. The last term of (%
\ref{Lcl})\ can be removed by translating $\phi $ and redefining the other
parameters, so we ignore it from now on. The theory is superrenormalizable
and the particle $\phi $ gets a nonvanishing width.

The quantization of $\varphi $ proceeds as usual, so we concentrate on $\phi 
$, starting from the free-field limit. The presence of $\varphi $ lets us
study the effects of interactions.

The $\phi $ momentum and its commutation relations read%
\begin{equation}
\pi _{\phi }=\rho \partial _{0}\phi ,\qquad \lbrack \pi _{\phi }(t,\mathbf{x}%
),\phi (t,\mathbf{y})]=-i\delta ^{(3)}(\mathbf{x}-\mathbf{y}).  \label{commo}
\end{equation}%
The free classical Hamiltonian is%
\begin{equation}
H_{\phi }=\frac{\rho }{2}\int \mathrm{d}^{3}\mathbf{x}\hspace{0.01in}\left[
(\partial _{0}\phi )^{2}+(\mathbf{\bigtriangledown }\phi )^{2}+M^{2}\phi ^{2}%
\right] .  \label{H}
\end{equation}%
To study more possibilities at once, we expand the field operator $\hat{\phi}
$ as%
\begin{equation}
\hat{\phi}(t,\mathbf{x})=\int \frac{\mathrm{d}^{3}\mathbf{k}}{(2\pi
)^{3}2\omega }\left[ (-1)^{\eta }a_{\mathbf{k}}\mathrm{e}^{-i\sigma
kx}+(-1)^{\eta ^{\prime }}a_{\mathbf{k}}^{\dag }\mathrm{e}^{i\sigma kx}%
\right] ,\qquad  \label{phi}
\end{equation}%
in terms of creation and annihilation operators $a_{\mathbf{k}}^{\dag }$ and 
$a_{\mathbf{k}}$, where $kx=\omega t-\mathbf{k}\cdot \mathbf{x}$ and $\omega
=\sqrt{\mathbf{k}^{2}+M^{2}}$. The parameters $\eta $ and $\eta ^{\prime }$
can have values 0 or 1, while $\sigma $ can have values $\pm 1$. To have
agreement with (\ref{commo}), the commutation relations of $a_{\mathbf{k}%
}^{\dag }$ and $a_{\mathbf{k}}$ must be 
\begin{equation}
\lbrack a_{\mathbf{k}},a_{\mathbf{k}^{\prime }}^{\dag }]=2\rho \sigma
(-1)^{\eta +\eta ^{\prime }}\omega (2\pi )^{3}\delta ^{(3)}(\mathbf{k}-%
\mathbf{k}^{\prime }),\qquad \lbrack a_{\mathbf{k}}^{\dag },a_{\mathbf{k}%
^{\prime }}^{\dag }]=[a_{\mathbf{k}},a_{\mathbf{k}^{\prime }}]=0.
\label{comma}
\end{equation}

We define the vacuum $|0\rangle $ to be annihilated by $a_{\mathbf{k}}$ and
the states to be created by $a_{\mathbf{k}}^{\dag }$:%
\begin{equation}
a_{\mathbf{k}}|0\rangle =0,\qquad |n\rangle =\frac{1}{\sqrt{n!A}}\int \left(
\prod\limits_{i=1}^{n}\frac{\mathrm{d}^{3}\mathbf{k}_{i}}{(2\pi )^{3}2\omega
_{i}}\right) f(\mathbf{k}_{1},\cdots ,\mathbf{k}_{n})a_{\mathbf{k}%
_{1}}^{\dagger }\cdots a_{\mathbf{k}_{n}}^{\dagger }|0\rangle ,
\label{states}
\end{equation}%
where%
\begin{equation*}
A=\int \left( \prod\limits_{i=1}^{n}\frac{\mathrm{d}^{3}\mathbf{k}_{i}}{%
(2\pi )^{3}2\omega _{i}}\right) |f(\mathbf{k}_{1},\cdots ,\mathbf{k}%
_{n})|^{2}.
\end{equation*}

From (\ref{H}) we derive the Hamiltonian operator%
\begin{equation}
\hat{H}_{\phi }=\frac{\rho }{2}(-1)^{\eta +\eta ^{\prime }}\int \frac{%
\mathrm{d}^{3}\mathbf{k}}{(2\pi )^{3}}a_{\mathbf{k}}^{\dagger }a_{\mathbf{k}}
\label{Ham}
\end{equation}%
(neglecting an infinite additive constant). From (\ref{comma}) we find the
norms%
\begin{equation}
\langle n|n\rangle =\rho ^{n}\sigma ^{n}(-1)^{n(\eta +\eta ^{\prime })}.
\label{norms}
\end{equation}%
The $\hat{H}_{\phi }$ eigenvalues are 
\begin{equation}
\hat{H}_{\phi }|n\rangle =h_{n}|n\rangle ,\qquad h_{n}=\sigma \sum_{i=1}^{n}%
\bar{\omega}_{i},  \label{eigen}
\end{equation}%
for $f(\mathbf{k}_{1},\cdots ,\mathbf{k}_{n})=\prod\limits_{i=1}^{n}\delta
^{(3)}(\mathbf{k}_{i}-\mathbf{\bar{k}}_{i})$ ($A$ being replaced by an
arbitrary finite constant), where $\mathbf{\bar{k}}_{i}$ are given momenta
and $\bar{\omega}_{i}$ are their frequencies.

The free ($T$ ordered) $\hat{\phi}$ propagator is%
\begin{eqnarray}
\langle 0|T\hat{\phi}(x)\hspace{0.01in}\hat{\phi}(y)|0\rangle &=&\rho \sigma
\int \frac{\mathrm{d}^{3}\mathbf{k}}{(2\pi )^{3}2\omega }\left[ \theta
(x^{0}-y^{0\hspace{0in}})\mathrm{e}^{-i\sigma k(x-y)}+\theta (y^{0\hspace{0in%
}}-x^{0})\mathrm{e}^{-i\sigma k(y-x)}\right]  \notag \\
&=&\int \frac{\mathrm{d}^{4}p}{(2\pi )^{4}}\frac{i\rho \mathrm{e}^{-ip(x-y)}%
}{p^{2}-M^{2}+i\sigma \epsilon }.  \label{propa}
\end{eqnarray}

The physical particles have $\rho =\sigma =1$ and $\eta =\eta ^{\prime }=0$.
Then the Hamiltonians $H_{\phi }$ and $\hat{H}_{\phi }$ are bounded from
below and the norms are positive.

Now we consider the options with $\rho =-1$.

\subsection{\texorpdfstring{$i\protect\epsilon $}{ie} ghost}

The first possibility is to perform the $\phi $ quantization as usual, which
means choose (\ref{phi}) with $\sigma =1$, $\eta =\eta ^{\prime }=0$. Then,
formula (\ref{norms}) shows that there are states with positive norms and
states with negative norms. From (\ref{eigen}), we see that the Hamiltonian $%
\hat{H}_{\phi }$ is bounded from below. Formula (\ref{propa}) shows that the
propagator is equal to (\ref{iep}), that is to say, the opposite of a
physical particle. In particular, the $i\epsilon $ prescription is the usual
one ($M^{2}\rightarrow M^{2}-i\epsilon $). This is just the ordinary ghost,
which has positive energy, but indefinite metric. We call it
\textquotedblleft $i\epsilon $ ghost\textquotedblright.

The $i\epsilon $ ghosts violate unitarity. Nevertheless, they satisfy a
pseudounitary equation (see \cite{diagrammar,diagrammatica}), which holds
perturbatively diagram by diagram.

Since we are assuming $M>2m$, the interaction equips $\phi $ with a positive
width $\Gamma _{\phi }$. The $\phi $ dressed propagator formally reads%
\begin{equation}
-\frac{iZ}{p^{2}-M_{\text{gh}}^{2}+i(\epsilon -M_{\text{gh}}\Gamma _{\phi })}
\label{dre1}
\end{equation}%
around the peak. The minus sign between $\epsilon $ and $M_{\text{gh}}\Gamma
_{\phi }$ signals that the resummation cannot be trusted close to the peak,
as shown in \cite{FselfE}, so we have a \textquotedblleft peak
uncertainty\textquotedblright . We cannot apply Veltman's projection,
because we do not know what dressed propagator we should use inside bigger
diagrams.

On the other hand, we cannot remove the ghosts from the external states
order by order in the perturbative expansion, since this kind of removal is
not a projection, because it is not compatible with the pseudounitarity
equation. Without projections, the classical limit is just (\ref{Lcl}).
Formula (\ref{H}) shows that the free $\phi $ classical Hamiltonian $H_{\phi
}$ is not bounded from below, although, as we have seen above, the quantum
one $\hat{H}_{\phi }$ is.

A way to overcome these obstacles is to \textit{define }the dressed
propagator of a ghost as the one of formula (\ref{dre1}) at $\epsilon =0$
and start over from there. Then we obtain the same options as with the $%
-i\epsilon $ ghost discussed below.

\subsection{\texorpdfstring{$-i\protect\epsilon $}{-ie} ghost}

Now we define $\phi $ in (\ref{phi}) with $\sigma =-1$ and $\eta =\eta
^{\prime }=0$. These choices give positive norms in (\ref{norms}), but
formula (\ref{eigen}) shows that the quantum Hamiltonian $\hat{H}_{\phi }$
in not bounded from below. The propagator (\ref{propa}) becomes (\ref{-iep})
and acquires an unusual prescription ($M^{2}\rightarrow M^{2}+i\epsilon $).

This option is not equivalent to the previous one, because the roles of the
annihilation and creation operators are interchanged inside $\phi $, but not
in the definitions (\ref{states}) of vacuum state and occupied states.

The dressed propagator can be resummed straightforwardly, including the
region around the peak, where we find (\ref{dre2}). Since there is no peak
uncertainty, Veltman's projection can be applied. Once we remove $\phi $
from the set of asymptotic states, because of its finite lifetime, $\phi $
is no longer a degree of freedom in the classical limit. This means that it
is \textquotedblleft frozen\textquotedblright , integrated out by means of
its own propagator (calculated at $\hbar \rightarrow 0$).

The classical limit is obtained by collecting the tree diagrams. Veltman's
projection reduces the set of such diagrams to those that do not have $\phi $
external legs. The $\phi $ internal legs build nonlocal interactions among
the physical fields $\varphi $. At the end, the true classical Lagrangian $%
\mathcal{L}_{\text{cl}}^{\text{V}}$ is the projected version of (\ref{Lcl}),
obtained by integrating out $\phi $ with the ghost propagator (\ref{-iep}),
which is the classical limit of (\ref{dre2}). The result is 
\begin{equation}
\mathcal{L}_{\text{cl}}^{\text{V}}=\frac{1}{2}(\partial _{\mu }\varphi
)(\partial ^{\mu }\varphi )-\frac{m^{2}}{2}\varphi ^{2}-\frac{g^{2}}{2}%
\varphi ^{2}\frac{1}{\square +M^{2}+i\epsilon }\varphi ^{2}.  \label{none}
\end{equation}%
As predicted, it contains a micro nonlocal $\varphi $ self-interaction,
which is also micro acausal. If $M$ is large enough to have agreement with
the experimental data available today, micro nonlocalities and micro
acausalities are not problematic. What makes $\mathcal{L}_{\text{cl}}^{\text{%
V}}$ not acceptable is that it is not Hermitian. The imaginary part of the
projected classical action is%
\begin{equation*}
\int \mathrm{d}^{4}x\hspace{0.01in}\text{Im}\left[ \mathcal{L}_{\text{cl}}^{%
\text{V}}\right] =\frac{\pi g^{2}}{2}\int \mathrm{d}^{4}x\hspace{0.01in}%
\varphi ^{2}\delta (-\square -M^{2})\varphi ^{2}=\frac{\pi g^{2}}{2}\int 
\frac{\mathrm{d}^{4}p}{(2\pi )^{4}}\hspace{0.01in}\widetilde{\varphi ^{2}}%
(-p)\delta (p^{2}-M^{2})\widetilde{\varphi ^{2}}(p),
\end{equation*}%
where $\widetilde{\varphi ^{2}}$ is the Fourier transform of $\varphi ^{2}$.

If we choose not to advocate Veltman's projection, the classical Lagrangian
we obtain is (\ref{Lcl}): the denominator $\square +M^{2}+i\epsilon $ is
moved to the numerator, sandwiched in between two fields $\phi $, so the
contribution of $-i\epsilon $ becomes negligible. In that case, $\phi $ is
not integrated out, but an independent degree of freedom, with its own field
equations and boundary conditions. The classical limit is Hermitian, but
still unacceptable, in the realm of perturbation theory, because the
classical free Hamiltonian $H_{\phi }$ is not bounded from below.

The Lee-Wick ghosts are $-i\epsilon $ ghosts equipped with appropriate rules
to treat them inside the loop diagrams (see section \ref{loops}).

\subsection{Non-Hermitian ghost}

We mention a third option, because it is the one preferred by Lee and Wick
in their papers, although it is equivalent to the $-i\epsilon $ ghost just
described. We choose $\rho =1$ and expand $\phi $ with $\sigma =1$, $\eta =0$%
, $\eta ^{\prime }=1$. So doing, we understand that $\phi $ is
anti-Hermitian and the coupling $g$ is purely imaginary. The metric is
indefinite and the Hamiltonians $H_{\phi }$ and $\hat{H}_{\phi }$ are
bounded from below. The dressed propagator is fine, so there is no peak
uncertainty and Veltman's projection can be applied. The $-i\epsilon $ ghost
can be obtained from this type of ghost, which we call non-Hermitian (nH)
ghost, by turning $\phi $ into $i\phi $ and $g$ into $-ig$.

There is also a variant with $\rho =1$, $\sigma =-1$, $\eta =0$, $\eta
^{\prime }=1$. Then, the norms are positive, but $\hat{H}_{\phi }$ is not
bounded from below. The dressed propagator cannot be resummed in the peak
region, so there is a peak uncertainty and Veltman's projection cannot be
used.

\subsection{Fakeon}

In the case of purely virtual particles, we can take $\rho =\pm 1$. Doubling
the set of creation and annihilation operators, we write%
\begin{equation*}
\hat{\phi}(t,\mathbf{x})=\int \frac{\mathrm{d}^{3}\mathbf{k}}{(2\pi
)^{3}2\omega }\left[ \frac{a_{\mathbf{k}}+b_{\mathbf{k}}^{\dag }}{\sqrt{2}}%
\mathrm{e}^{-ikx}+\frac{a_{\mathbf{k}}^{\dag }+b_{\mathbf{k}}}{\sqrt{2}}%
\mathrm{e}^{ikx}\right] \qquad
\end{equation*}%
and assume%
\begin{equation*}
\lbrack a_{\mathbf{k}},a_{\mathbf{k}^{\prime }}^{\dag }]=2\rho \omega (2\pi
)^{3}\delta ^{(3)}(\mathbf{k}-\mathbf{k}^{\prime }),\qquad \lbrack b_{%
\mathbf{k}},b_{\mathbf{k}^{\prime }}^{\dag }]=-2\rho \omega (2\pi
)^{3}\delta ^{(3)}(\mathbf{k}-\mathbf{k}^{\prime }),
\end{equation*}%
all the other commutators being identically zero.

Inside the loop diagrams, the fakeon projection amounts to integrate $\phi $
out with the appropriate diagrammatic rules (see \cite{diagrammarMio} for
explicit formulas). In the classical limit, we must integrate it out with
the propagator\ 
\begin{equation}
\mathcal{P}\frac{i\rho }{p^{2}-M^{2}},  \label{Cauchy}
\end{equation}%
which coincides with the Fourier transform of $\langle 0|T\hat{\phi}(x)%
\hspace{0.01in}\hat{\phi}(y)|0\rangle $, where $\mathcal{P}$ is the Cauchy
principal value. The Lagrangian describing the classical limit, which reads 
\begin{equation}
\mathcal{L}_{\text{cl}}^{\text{f}}=\frac{1}{2}(\partial _{\mu }\varphi
)(\partial ^{\mu }\varphi )-\frac{m^{2}}{2}\varphi ^{2}+\rho \frac{g^{2}}{2}%
\varphi ^{2}\mathcal{P}\left( \frac{1}{\square +M^{2}}\right) \varphi ^{2},
\label{Lclf}
\end{equation}%
is the sum of a standard kinetic term plus a micro nonlocal Hermitian
self-interaction.

\subsection{Summary}

We summarize the various options considered so far and their main properties
in table \ref{table1}, where \textquotedblleft l\textquotedblright\ stands
for local, \textquotedblleft nl\textquotedblright\ means nonlocal,
\textquotedblleft +V\textquotedblright\ and \textquotedblleft
--V\textquotedblright\ mean with\ and without Veltman's projection,
respectively, and \textquotedblleft f$^{\pm }$\textquotedblright\ denotes
the fakeons with $\rho =\pm 1$. Finally, \textquotedblleft phys.
part.\textquotedblright\ means physical particle, \textquotedblleft
uncert.\textquotedblright\ means uncertainty, \textquotedblleft
Re\textquotedblright\ means Hermitian and \textquotedblleft
Im\textquotedblright\ means non-Hermitian.

\begin{table}[t]
\begin{center}
\begin{tabular}{|c|c|c|c|c|c|c|c|c|}
\hline
& 
\begin{tabular}{c}
{\footnotesize phys.} \\ 
{\footnotesize part.}%
\end{tabular}
& $i\epsilon \hspace{0.01in}${\footnotesize gh} & $-i\epsilon \hspace{0.01in}
${\footnotesize gh+V} & $-i\epsilon \hspace{0.01in}${\footnotesize gh--V} & 
{\footnotesize nH+V} & {\footnotesize nH--V} & {\footnotesize nH}$^{\sigma
=-1}$ & {\footnotesize f}$^{\pm }$ \\ \hline
$\rho $ & $1$ & $-1$ & $-1$ & $-1$ & $1$ & $1$ & $1$ & $\pm 1$ \\ \hline
$\sigma $ & $1$ & $1$ & $-1$ & $-1$ & $1$ & $1$ & $-1$ & $\pm 1$ \\ \hline
$\eta $ & $0$ & $0$ & $0$ & $0$ & $0$ & $0$ & $0$ & $0$ \\ \hline
$\eta ^{\prime }$ & $0$ & $0$ & $0$ & $0$ & $1$ & $1$ & $1$ & $0$ \\ \hline
$\hat{\phi}^{\dagger }$ & $\hat{\phi}$ & $\hat{\phi}$ & $\hat{\phi}$ & $\hat{%
\phi}$ & $-\hat{\phi}$ & $-\hat{\phi}$ & $-\hat{\phi}$ & $\hat{\phi}$ \\ 
\hline
{\footnotesize norms} & $+$ & $\pm $ & $\times $ & $+$ & $\times $ & $\pm $
& $+$ & $\times $ \\ \hline
$H_{\phi }$ & $\geqslant 0$ & $\leqslant 0$ & $\times $ & $\leqslant 0$ & $%
\times $ & $\geqslant 0$ & $\geqslant 0$ & $\times $ \\ \hline
$\hat{H}_{\phi }$ & $\geqslant 0$ & $\geqslant 0$ & $\times $ & $\leqslant 0$
& $\times $ & $\geqslant 0$ & $\leqslant 0$ & $\times $ \\ \hline
$\mathcal{L}_{\text{cl}}$ & 
\begin{tabular}{c}
{\footnotesize l} \\ 
{\footnotesize Re}%
\end{tabular}
& 
\begin{tabular}{c}
{\footnotesize l} \\ 
{\footnotesize Re}%
\end{tabular}
& 
\begin{tabular}{c}
{\footnotesize nl} \\ 
{\footnotesize Im}%
\end{tabular}
& 
\begin{tabular}{c}
{\footnotesize l} \\ 
{\footnotesize Re}%
\end{tabular}
& 
\begin{tabular}{c}
{\footnotesize nl} \\ 
{\footnotesize Im}%
\end{tabular}
& 
\begin{tabular}{c}
{\footnotesize l} \\ 
{\footnotesize Re}%
\end{tabular}
& 
\begin{tabular}{c}
{\footnotesize l} \\ 
{\footnotesize Re}%
\end{tabular}
& 
\begin{tabular}{c}
{\footnotesize nl} \\ 
{\footnotesize Re}%
\end{tabular}
\\ \hline
\begin{tabular}{c}
{\footnotesize peak} \\ 
{\footnotesize uncert.}%
\end{tabular}
& no & $\checkmark $ & no & no & no & no & $\checkmark $ & $\checkmark $ \\ 
\hline
\end{tabular}%
\end{center}
\caption{Main properties of the options for quantization. The symbol
\textquotedblleft $\times $\textquotedblright\ means \textquotedblleft not
applicable\textquotedblright }
\label{table1}
\end{table}

\section{Loops and unstable particles}

\label{loops}\setcounter{equation}{0}

We briefly recall how the various options listed in the previous section are
treated inside the loop diagrams, referring to the literature for more
details.

The propagator of an $i\epsilon $ ghost coincides with the one of a physical
particle, apart from its overall sign, so its diagrammatics is
straightforward. The propagator of a $-i\epsilon $ ghost, on the other hand,
is defined by the opposite prescription. If we integrate the loop diagrams
on real energies and real momenta, the $-i\epsilon $ prescription cannot
coexist with the usual $i\epsilon $ one \cite{wheelerons}, because it
switches the roles of the thresholds with those of the pseudothresholds,
causing instabilities, violations of unitarity, as well as violations of the
locality and Hermiticity of counterterms. To avoid these types of problems,
it is necessary to formulate better integration prescriptions or give
alternative diagrammatic rules.

The LW ghost is obtained from the $-i\epsilon $ ghost by adopting the
Lee-Wick integration prescription on the loop energies \cite{leewick},
combined with the CLOP prescription \cite{CLOP} and any other rules that
might be necessary for the internal consistency. Here, we do not need to
prove that they exist, so we just assume that they do. The LW ghosts must be
unstable, dynamically or not, to apply Veltman's projection to them. The
procedure is semiperturbative, because it requires to use the dressed
propagators inside diagrams and reorganize the diagrammatic rules
accordingly.

In the model (\ref{Lcl}), it is enough to turn on the vertex $-g\varphi
^{2}\phi $ and assume the inequality $M>2m$. Then, the decay $\phi
\rightarrow \varphi \varphi $ gives $\phi $ a nonvanishing width $\Gamma
_{\phi }$. The $-i\epsilon $ prescription guarantees that it is possible to
resum the self-energies into the $\phi $ dressed propagators with no peak
uncertainty, so Veltman's projection can be applied to $\phi $, to build a
unitary reduced $S$ matrix $S_{r}$ on the strictly asymptotic states.

Fakeons inside loops are defined by means of a different diagrammatics \cite%
{diagrammarMio}, which works as a mathematical tool to surgically eradicate
the potential degree of freedom at all energies, turning it \textit{de facto}
into a fake degree of freedom. The mathematics of the fakeon projection does
not have a direct physical interpretation, such as a decay. We may expect
that if the removal of a degree of freedom (or its impossibility to be
observed in nature) is due to the physics, it is either nonperturbative (as
in the cases of quarks and gluons) or not a complete removal (as in the case
of the LW ghost).

In particular, a purely virtual particle does not need to have a
nonvanishing width $\Gamma $. The assumption $M>2m$ is unnecessary to make
the model (\ref{Lcl}) work with $\phi $ = fakeon. We can even replace the
vertex $-g\varphi ^{2}\phi $ with an interaction like $-g\varphi \phi ^{2}$,
which makes the fakeon width identically zero. Phenomenological models with
fakeons of vanishing widths are studied in ref. \cite{Tallinn1}.

\section{Pauli-Villars fields made physical}

\label{PV}\setcounter{equation}{0}

In this section we use fakeons to make the Pauli-Villars fields consistent
and observable without sending their masses to infinity.

We first recall the main properties of the Pauli-Villars fields \cite%
{PV,diagrammar}. Consider the Lagrangian%
\begin{equation}
\mathcal{L}_{\text{PV}}=\frac{1}{2}\sum_{j=1}^{N}\left[ (\partial _{\mu
}\varphi _{j})(\partial ^{\mu }\varphi _{j})-m_{j}^{2}\varphi _{j}^{2}\right]
-\frac{1}{2}\sum_{j=1}^{N^{\prime }}\left[ (\partial _{\mu }\phi
_{j})(\partial ^{\mu }\phi _{j})-M_{j}^{2}\phi _{j}^{2}\right] -V\left( 
\breve{\varphi},\phi \right) ,  \label{LPV}
\end{equation}%
where%
\begin{equation*}
\phi \equiv \sum_{j=1}^{N}c_{j}\varphi _{j}+\sum_{j=1}^{N^{\prime
}}d_{j}\phi _{j},
\end{equation*}%
$c_{j}$, $d_{j}$ are real constants and $\breve{\varphi}$ are the fields $%
\varphi _{j}$ that do not appear inside $\phi $ (because they have $c_{j}=0$%
). $V$ is a potential, or, more generally, the interaction part (if it
depends on the derivatives of the fields).

It is possible to organize the diagrammatics so that each non-$\breve{\varphi%
}$ internal leg of the diagrams propagates the whole combination $\phi $,
with free propagator%
\begin{equation}
\langle \phi (p)\hspace{0.01in}\phi (-p)\rangle _{0}=\sum_{j=1}^{N}\frac{%
ic_{j}^{2}}{p^{2}-m_{j}^{2}+i\epsilon }-\sum_{j=1}^{N^{\prime }}\frac{%
id_{j}^{2}}{p^{2}-M_{j}^{2}+i\epsilon }.  \label{cprop}
\end{equation}%
For the moment, we use the standard $i\epsilon $ prescription for the PV
fields $\phi _{j}$. We examine different options later.

If we choose $c_{j}$, $d_{j}$ such that%
\begin{equation}
\sum_{j=1}^{N}c_{j}^{2}(m_{j}^{2})^{k}=\sum_{j=1}^{N^{\prime
}}d_{j}^{2}(M_{j}^{2})^{k},  \label{cpv2}
\end{equation}%
$k=0,1,2,\ldots \bar{n}$, the propagator (\ref{cprop}) falls off as $%
1/(p^{2})^{2+\bar{n}}$ for large $|p^{2}|$. So doing, we can improve the
power counting and in some cases render the theory (\ref{LPV}) completely
finite.

If we plan to send the masses $M_{j}$ to infinity, we can use the PV\ fields
as regulators. The Pauli-Villars regularization technique is obtained by
adding PV\ fields so as to make the theory completely finite at finite
masses $M_{j}$.

An important property of the PV regularization technique is that it is not
gauge invariant, nor general covariant, because it does not treat the
quadratic and interaction parts of the Lagrangian on an equal footing.
Indeed, the interaction part of (\ref{LPV}) depends only on the physical
fields $\breve{\varphi}$ and the linear combination $\phi $, while the
quadratic terms cannot be expressed by means of $\breve{\varphi}$ and $\phi $%
. Gauge invariance and general covariance can be recovered in the limit $%
M_{j}\rightarrow \infty $ (provided they are not anomalous) by subtracting
local counterterms.

If we want to give physical significance to the PV fields without sending
their masses to infinity, we must restrict to neutral matter fields. We
study the main options we have in this context.

The first option is the one already considered, i.e., quantize the PV fields
as $i\epsilon $ ghosts. Then, it is not possible to ignore them from the
incoming\ and outgoing\ states, because the dressed propagators cannot be
resummed around the peaks. The classical Lagrangian is not acceptable,
because it has negative kinetic terms.

The second option is what Lee and Wick do, i.e., quantize the PV fields as $%
-i\epsilon $ ghosts, treat them as LW ghosts inside the loop diagrams,
ensure that they are unstable, build their dressed propagators and apply
Veltman's projection. The problem with this option is that the classical
limit is not Hermitian.

The standard option with fakeons is to quantize $\phi _{j}$ as purely
virtual particles, which removes them completely. There is a new
possibility, though, which emerges by combining the PV, LW and fakeon ideas
in a certain way. It amounts to removing the $\phi _{j}$ only partially, to
overcome the difficulties described above and gain extra (observable)
physical particles. The trick works with neutral matter fields and is fully
perturbative.

We first describe the new option in the model (\ref{Lcl}), with $\Lambda =0$%
, which is a particular case of (\ref{LPV}), then we generalize it to (\ref%
{LPV}). In the next sections we apply it to finite QED,\ and quantum gravity.

We decompose $\phi $ as the combination of a physical field $\Phi $ and an
additional field $Q$. Specifically, we turn the classical Lagrangian (\ref%
{Lcl}) into%
\begin{eqnarray}
\mathcal{L}_{\text{cl}} &=&\frac{1}{2}(\partial _{\mu }\varphi )(\partial
^{\mu }\varphi )-\frac{m^{2}}{2}\varphi ^{2}+\frac{1}{2}(\partial _{\mu
}\Phi )(\partial ^{\mu }\Phi )-\frac{M^{2}}{2}\Phi ^{2}  \notag \\
&&-\frac{1}{2}\left[ (\partial _{\mu }Q)(\partial ^{\mu }Q)-M^{2}Q^{2}\right]
-g\varphi ^{2}(\Phi +\sqrt{2}Q)  \label{repl}
\end{eqnarray}%
and view $\Phi $ as a standard physical particle and $Q$ as a fakeon. The
combined $\phi $ propagator becomes%
\begin{equation}
-\left. \frac{2i}{p^{2}-M^{2}}\right\vert _{\text{f}}+\frac{i}{%
p^{2}-M^{2}+i\epsilon },  \label{pp2}
\end{equation}%
where the subscript \textquotedblleft f\textquotedblright\ means
\textquotedblleft fakeon prescription\textquotedblright . At the tree level,
(\ref{Cauchy}) (with $\rho \rightarrow -2$) gives 
\begin{equation*}
-\frac{i}{p^{2}-M^{2}-i\epsilon }-\frac{i}{p^{2}-M^{2}+i\epsilon }+\frac{i}{%
p^{2}-M^{2}+i\epsilon }=-\frac{i}{p^{2}-M^{2}-i\epsilon },
\end{equation*}%
which is the propagator of a $-i\epsilon $ ghost. Resumming the
self-energies, the dressed propagator reads%
\begin{equation*}
-\frac{iZ}{p^{2}-M_{\text{ph}}^{2}-i(\epsilon +\Gamma )}\rightarrow -\frac{iZ%
}{p^{2}-M_{\text{ph}}^{2}-i\Gamma }
\end{equation*}%
around the peak, where $\Gamma $ is non-negative. Although the $\phi $
two-point function has no peak uncertainty, this is not crucial now, because
we do not need Veltman's projection, having abandoned the LW\ approach to
adopt the fakeon one.

Since $\Phi $ is a fakeon, it does not belong to the set of asymptotic
states, by definition. Instead, $Q$ does: unstable or not, it is an extra,
physically observable particle, originated by a PV field. The decomposition
of $\phi $ as $\Phi +\sqrt{2}Q$ and the properties of fakeons allow us to
treat $\Phi $ and $Q$ differently, which is crucial to have a Hermitian
classical limit. That limit, obtained by keeping $\Phi $ and projecting $Q$
away in (\ref{repl}), is given by the Lagrangian%
\begin{equation}
\mathcal{L}_{\text{cl}}^{\text{f}}=\frac{1}{2}(\partial _{\mu }\varphi
)(\partial ^{\mu }\varphi )-\frac{m^{2}}{2}\varphi ^{2}+\frac{1}{2}(\partial
_{\mu }\Phi )(\partial ^{\mu }\Phi )-\frac{M^{2}}{2}\Phi ^{2}-g\varphi
^{2}\Phi -g^{2}\varphi ^{2}\mathcal{P}\frac{1}{\square +M^{2}}\varphi ^{2},
\label{lclf}
\end{equation}%
which describes two physical particles, $\varphi $ and $\Phi $, with a
nonstandard (micro nonlocal and micro acausal) $\varphi $ self-interaction.

The decomposition also specifies how to treat $\phi $ inside the loop
diagrams. We must proceed as in \cite{diagrammarMio}, distinguishing the
contributions due to $\Phi $ from those due to $Q$, since $\Phi $ is a
physical particle, while $Q$ is a fakeon. Note that $\Phi $ and $Q$ have the
same mass, so there are many coinciding thresholds, which must be treated as
limits of distinct ones. The counterterms of (\ref{repl}) just depend on $%
\varphi $ and $\phi $. The $\phi $ propagator renormalizes exactly as for
the theory (\ref{Lcl}). The $\Phi $ and $Q$ two-point functions can be
derived from it. They may separately have peak uncertainties, but, again,
this is not of our concern.

The masses of purely virtual particles are observable quantities.
Nevertheless, they are not revealed as \textquotedblleft
masses\textquotedblright , but through their indirect effects on the other
particles. For example, in the model (\ref{lclf}) such effects are encoded
in the last term, $M$ being the mass of the fakeon $Q$. %Since $M$ is also
%the mass of $\Phi $, which is a physical particle, it can be seen as a
%physical mass of $\Phi $. The coincidence of the two is a prediction of the
%model.

We can generalize the trick by turning (\ref{LPV}) into%
\begin{eqnarray}
\mathcal{L}_{\text{PV}} &=&\frac{1}{2}\sum_{j=1}^{N}\left[ (\partial _{\mu
}\varphi _{j})(\partial ^{\mu }\varphi _{j})-m_{j}^{2}\varphi _{j}^{2}\right]
+\frac{1}{2}\sum_{j=1}^{N^{\prime }}\left[ (\partial _{\mu }\Phi
_{j})(\partial ^{\mu }\Phi _{j})-M_{j}^{2}\Phi _{j}^{2}\right]  \notag \\
&&-\frac{1}{2}\sum_{j=1}^{N^{\prime }}\left[ (\partial _{\mu
}Q_{j})(\partial ^{\mu }Q_{j})-M_{j}^{2}Q_{j}^{2}\right] -V\left( \check{%
\varphi},\phi \right) ,  \label{repl2}
\end{eqnarray}%
where%
\begin{equation*}
\phi =\sum_{j=1}^{N}c_{j}\varphi _{j}+\sum_{j=1}^{N^{\prime }}d_{j}(\Phi
_{j}+\sqrt{2}Q_{j}),
\end{equation*}%
and interpreting $\Phi _{j}$ as additional physical particles and $Q_{j}$ as
fakeons. The $\phi $ propagator is 
\begin{equation*}
\sum_{j=1}^{N}\frac{ic_{j}^{2}}{p^{2}-m_{j}^{2}+i\epsilon }%
+\sum_{j=1}^{N^{\prime }}\frac{id_{j}^{2}}{p^{2}-M_{j}^{2}+i\epsilon }%
-\sum_{j=1}^{N^{\prime }}\left. \frac{2id_{j}^{2}}{p^{2}-M_{j}^{2}}%
\right\vert _{\text{f}}.
\end{equation*}%
Thanks to conditions such as (\ref{cpv2}), we can make it fall off as fast
as we want for large $|p^{2}|$. The classical limit is a theory of $%
N+N^{\prime }$ physical particles with certain micro nonlocal
self-interactions.

\section{Finite QED}

\label{QED}\setcounter{equation}{0}

In this section we use the trick explained in the previous one to build a
finite QED, by tweaking the original Lee-Wick construction \cite{LWQED}. In
the next section we investigate the possibility of generalizing the trick to
quantum gravity.

We start from the classical Lagrangian%
\begin{eqnarray}
\mathcal{L}_{\text{QED}} &=&-\frac{1}{4}F_{\mu \nu }F^{\mu \nu }+\frac{1}{4}%
G_{\mu \nu }G^{\mu \nu }-\frac{M^{2}}{2}B_{\mu }B^{\mu }+\sum_{j=1}^{2}\bar{%
\psi}_{j}\left[ i\gamma ^{\mu }(\partial _{\mu }+ieA_{\mu }+ieB_{\mu })-m_{j}%
\right] \psi _{j}  \notag \\
&&+\bar{\Psi}\sigma _{1}\left[ i\gamma ^{\mu }(\partial _{\mu }+e\sigma
_{2}A_{\mu }+e\sigma _{2}B_{\mu })-M_{\Psi }\right] \Psi ,  \label{LQED}
\end{eqnarray}%
where $F_{\mu \nu }=\partial _{\mu }A_{\nu }-\partial _{\nu }A_{\mu }$, $%
G_{\mu \nu }=\partial _{\mu }B_{\nu }-\partial _{\nu }B_{\mu }$, $\sigma
_{1} $ and $\sigma _{2}$ are the first two Pauli matrices, $\psi _{1}$ and $%
\psi _{2}$ denote the electron and the muon, respectively, $\Psi $ is an
extra fermion doublet and $M_{\Psi }$ denotes the $\Psi $ mass matrix. The
Lagrangian is Hermitian and gauge invariant, the gauge transformations being%
\begin{equation*}
A_{\mu }\rightarrow A_{\mu }+\partial _{\mu }\Lambda ,\quad B_{\mu
}\rightarrow B_{\mu },\quad \psi \rightarrow \mathrm{e}^{-ie\Lambda }\psi
,\quad \bar{\psi}\rightarrow \mathrm{e}^{ie\Lambda }\bar{\psi},\quad \Psi
\rightarrow \mathrm{e}^{-e\sigma _{2}\Lambda }\Psi ,\quad \bar{\Psi}%
\rightarrow \bar{\Psi}\mathrm{e}^{-e\sigma _{2}\Lambda }.
\end{equation*}

\subsection{Lee-Wick QED}

If we follow Lee and Wick, the vector $B_{\mu }$ is a $-i\epsilon $ ghost at
the tree level, to be treated as a LW ghost inside the loops. Since the
interactions contain only the combination $A_{\mu }+B_{\mu }$, what matters,
for power counting, is the combined propagator of $A_{\mu }+B_{\mu }$, which
reads%
\begin{equation}
-i\eta _{\mu \nu }\left( \frac{1}{p^{2}+i\epsilon }-\frac{1}{%
p^{2}-M^{2}-i\epsilon }\right) +p_{\mu }p_{\nu }(\cdots ),  \label{comba}
\end{equation}%
where $\eta _{\mu \nu }=$diag$(1,-1,-1,-1)$ is the flat-space metric. The
transverse part, proportional to $\eta _{\mu \nu }$, is gauge independent
and falls off like $1/(p^{2})^{2}$ for large momenta, \`{a} la
Pauli-Villars. The longitudinal part does not fall off rapidly.
Nevertheless, it is gauge dependent and does not affect the physical
quantities.

The behavior of (\ref{comba}) is enough to ensure that every diagram but one
is convergent, up to gauge-dependent contributions. The exception is the
one-loop photon self-energy. Its convergence is provided by the doublet $%
\Psi $, introduced to obtain a completely finite theory.

At one loop the photon self-energy receives contributions from the bubble
diagrams with circulating electrons, muons and $\Psi $ fields. The diagram
with circulating $\Psi $ fields has an extra $-2$ factor with respect to the
electron and muon bubble diagrams, because of the trace%
\begin{equation*}
\text{tr}[\sigma _{1}(-i\sigma _{1}\sigma _{2})\sigma _{1}(-i\sigma
_{1}\sigma _{2})]=-2.
\end{equation*}%
A $\sigma _{1}$ is brought by each $\Psi $ propagator and a $-i\sigma
_{1}\sigma _{2}$ is brought by each vertex. The factor $-2$ is precisely
what is needed to compensate the logarithmic divergences due to electrons
and muons.

The dressed propagator of $A_{\mu }+B_{\mu }$ can be resummed in the
transverse sector. We do not repeat the calculation of Lee and Wick here,
but just recall that $B_{\mu }$ acquires a nonvanishing width and becomes
unstable. It is then removed from the set of strictly asymptotic states \`{a}
la Veltman.

Lee and Wick need to make $\Psi $ decay as well. Since $\Psi $ does not
become unstable dynamically, they equip it with a nonvanishing width at the
classical level, by choosing a mass matrix of the form%
\begin{equation}
M_{\Psi }=m_{\Psi }+\frac{i}{2}\sigma _{2}\gamma _{\Psi },  \label{massM}
\end{equation}%
where $m_{\Psi }$ and $\gamma _{\Psi }$ are real numbers. The Lagrangian (%
\ref{LQED}) remains Hermitian.

Once Veltman's projection is advocated for the unstable particles $B_{\mu }$
and $\Psi $ (the muon being stable here), the reduced $S$ matrix $S_{r}$ is
unitary. As expected, the classical limit, which reads%
\begin{eqnarray}
\mathcal{L}_{\text{cl}}^{\text{LW}} &=&-\frac{1}{4}F_{\mu \nu }F^{\mu \nu
}+\sum_{j=1}^{2}\bar{\psi}_{j}\left[ i\gamma ^{\mu }(\partial _{\mu
}+ieA_{\mu })-m_{j}\right] \psi _{j}  \notag \\
&&+\frac{e^{2}}{2M^{2}}\left( \sum_{j=1}^{2}\bar{\psi}_{j}\gamma ^{\mu }\psi
_{j}\right) \frac{\eta _{\mu \nu }M^{2}+\partial _{\mu }\partial _{\nu }}{%
\square +M^{2}+i\epsilon }\left( \sum_{l=1}^{2}\bar{\psi}_{l}\gamma ^{\nu
}\psi _{l}\right) ,  \label{LclQEDLW}
\end{eqnarray}%
contains a non-Hermitian self-interaction.

\subsection{Standard option with fakeons}

The easiest way to solve the problems of the Lee-Wick construction is to
treat $B_{\mu }$ and $\Psi $ as fakeons. The theory remains finite. The
dressed $B_{\mu }$ propagator cannot be resummed around its peak, so $B_{\mu
}$ has a peak uncertainty, equal to its width divided by 2 \cite{FselfE}. As
far as $\Psi $ is concerned, we can just leave $\gamma _{\Psi }=0$ in
formula (\ref{massM}), since $\Psi $ is out of the physical spectrum without
requiring that it decays. Note that $\Psi $ appears quadratically in the
action. This means that, once it is projected away, it does not contribute
to the classical limit (its field equation being satisfied by $\Psi =0$). At
higher orders, it contributes by means of $\Psi $ loops (similar to the
loops of Faddeev-Popov ghosts), which are Hermitian due to the diagrammatics
of purely virtual particles.

The classical limit becomes%
\begin{eqnarray}
\mathcal{L}_{\text{cl}}^{\text{f}} &=&-\frac{1}{4}F_{\mu \nu }F^{\mu \nu
}+\sum_{j=1}^{2}\bar{\psi}_{j}\left[ i\gamma ^{\mu }(\partial _{\mu
}+ieA_{\mu })-m_{j}\right] \psi _{j}  \notag \\
&&+\frac{e^{2}}{2M^{2}}\left( \sum_{j=1}^{2}\bar{\psi}_{j}\gamma ^{\mu }\psi
_{j}\right) \mathcal{P}\frac{\eta _{\mu \nu }M^{2}+\partial _{\mu }\partial
_{\nu }}{\square +M^{2}}\left( \sum_{l=1}^{2}\bar{\psi}_{l}\gamma ^{\nu
}\psi _{l}\right) ,  \label{LclQEDf}
\end{eqnarray}%
which is the standard QED\ Lagrangian with a nonstandard four fermion
Hermitian, micro nonlocal self-interaction.

\subsection{New option with fakeons}

The new option, instead, amounts to interpreting $B_{\mu }$ as a
superposition $\tilde{B}_{\mu }+\sqrt{2}Q_{\mu }$ of a physical vector $%
\tilde{B}_{\mu }$ and a different fakeon $Q_{\mu }$, while the doublet $\Psi 
$ is still seen as a fakeon. We obtain the Lagrangian 
\begin{eqnarray}
\mathcal{L}_{\text{QED}} &=&-\frac{1}{4}F_{\mu \nu }F^{\mu \nu }-\frac{1}{4}%
\tilde{F}_{\mu \nu }\tilde{F}^{\mu \nu }+\frac{M^{2}}{2}\tilde{B}_{\mu }%
\tilde{B}^{\mu }+\frac{1}{4}Q_{\mu \nu }Q^{\mu \nu }-\frac{M^{2}}{2}Q_{\mu
}Q^{\mu }  \notag \\
&&+\sum_{j=1}^{2}\bar{\psi}_{j}\left[ i\gamma ^{\mu }(\partial _{\mu
}+ieA_{\mu }+ie\tilde{B}_{\mu }+ie\sqrt{2}Q_{\mu })-m_{j}\right] \psi _{j} 
\notag \\
&&+\bar{\Psi}\sigma _{1}\left[ i\gamma ^{\mu }(\partial _{\mu }+e\sigma
_{2}A_{\mu }+e\sigma _{2}\tilde{B}_{\mu }+e\sqrt{2}\sigma _{2}Q_{\mu
})-M_{\Psi }\right] \Psi ,  \label{extaQED}
\end{eqnarray}%
where $\tilde{F}_{\mu \nu }=\partial _{\mu }\tilde{B}_{\nu }-\partial _{\nu }%
\tilde{B}_{\mu }$ and $Q_{\mu \nu }=\partial _{\mu }Q_{\nu }-\partial _{\nu
}Q_{\mu }$. The theory remains finite, because the combined propagator of $%
A_{\mu }+\tilde{B}_{\mu }+\sqrt{2}Q_{\mu }$ behaves like (\ref{comba}) for
large $|p^{2}|$, although it is defined by a different prescription at
finite momenta. The classical limit reads%
\begin{eqnarray}
\mathcal{L}_{\text{cl}}^{\text{f}} &=&-\frac{1}{4}F_{\mu \nu }F^{\mu \nu }-%
\frac{1}{4}\tilde{F}_{\mu \nu }\tilde{F}^{\mu \nu }+\frac{M^{2}}{2}\tilde{B}%
_{\mu }\tilde{B}^{\mu }+\sum_{j=1}^{2}\bar{\psi}_{j}\left[ i\gamma ^{\mu
}(\partial _{\mu }+ieA_{\mu }+ie\tilde{B}_{\mu })-m_{j}\right] \psi _{j} 
\notag \\
&&+\frac{e^{2}}{M^{2}}\left( \sum_{j=1}^{2}\bar{\psi}_{j}\gamma ^{\mu }\psi
_{j}\right) \mathcal{P}\frac{\eta _{\mu \nu }M^{2}+\partial _{\mu }\partial
_{\nu }}{\square +M^{2}}\left( \sum_{l=1}^{2}\bar{\psi}_{l}\gamma ^{\nu
}\psi _{l}\right) .
\end{eqnarray}%
This is the standard QED\ Lagrangian with an extra Proca vector $\tilde{B}%
_{\mu }$ and a peculiar Hermitian, micro nonlocal four fermion
self-interaction.

Since the QED\ models formulated in this section are finite, the coupling $%
\alpha =e^{2}/(4\pi )$ does not run. However, if the mass $M$ of the vector $%
B_{\mu }$ and the mass $M_{\Psi }$ of the doublet $\Psi $ are assumed to be
large, they can be treated as cutoffs at low energies. The logarithmic
divergences that appear when they tend to infinity give the usual running.

\section{Quantum gravity}

\label{QG}\setcounter{equation}{0}

In this section we discuss the possibility of applying the trick to quantum
gravity and stress the difficulties that arise with general covariance. Then
we explain how a fully covariant decomposition can be achieved by adding a
massive spin-2 multiplet.

Consider the classical action%
\begin{equation}
S_{\text{QG}}=-\frac{1}{16\pi G}\int \mathrm{d}^{4}x\sqrt{-g}\left( 2\Lambda
+R+\frac{\lambda }{2m_{\chi }^{2}}C_{\mu \nu \rho \sigma }C^{\mu \nu \rho
\sigma }-\frac{R^{2}}{6m_{\phi }^{2}}\right) ,  \label{sqg}
\end{equation}%
where $\lambda ={m_{\chi }^{2}(3m_{\phi }^{2}+4\Lambda )}/(m_{\phi
}^{2}(3m_{\chi }^{2}-2\Lambda ))$ is a parameter very close to 1. The theory
includes the square $C_{\mu \nu \rho \sigma }C^{\mu \nu \rho \sigma }$ of
the Weyl tensor $C_{\mu \nu \rho \sigma }$ and is renormalizable by power
counting \cite{stelle}. It propagates the graviton, a scalar field $\phi _{%
\text{infl}}$ of mass $m_{\phi }$ (which can be interpreted as the inflaton)
and a spin-2 field $\chi _{\mu \nu }$ of mass $m_{\chi }$, which has a
kinetic term multiplied by the wrong sign. The three can be made explicit
with the help of auxiliary fields, as shown in \cite{Absograv}.

If $\chi _{\mu \nu }$ is interpreted as an $i\epsilon $ ghost, we obtain the
Stelle theory \cite{stelle}, which is not unitary. Since $\chi _{\mu \nu }$
dynamically acquires a nonvanishing width $\Gamma _{\chi }$, it is
interesting to consider the $\chi _{\mu \nu }$ dressed propagator. The
results of \cite{FselfE} show that the resummation of the self-energies does
not make sense around the peak, because we formally obtain (\ref{dre1}).
Thus, we cannot project $\chi _{\mu \nu }$ away \`{a} la Veltman. The
classical limit of this theory is exactly (\ref{sqg}), which is not
acceptable.

The LW option, studied by Donoghue and Menezes in ref. \cite{donoghue}, is
to interpret $\chi _{\mu \nu }$ as a $-i\epsilon $ ghost at the tree level
and as a LW ghost inside the loops. In this case, it is meaningful to resum
the self-energies into the $\chi _{\mu \nu }$ dressed propagator, which has
the form (\ref{dre2}). It is possible to project $\chi _{\mu \nu }$ away 
\`{a} la Veltman and focus on the reduced $S$ matrix $S_{r}$. However, the
classical limit is not Hermitian, like (\ref{none}) and (\ref{LclQEDLW}).

If we tweak the Lee-Wick proposal by removing Veltman's projection, the
classical limit is still (\ref{sqg}). If we remove Veltman's projection just
for the physical massive spin-2 particle singled out by the narrow-width
approximation (the real part of the propagator (\ref{-iep})), we break
general covariance, for arguments similar to the ones we explain below.

Other Lee-Wick approaches to quantum gravity, starting from different
classical actions, have been considered in the literature \cite{otherLWgra}.

\bigskip

Now we examine the options we have with fakeons. The standard option is to
interpret $\chi _{\mu \nu }$ as a fakeon, which gives the quantum gravity
theory of \cite{LWgrav}. Then, $\chi _{\mu \nu }$ does not belong to the
sets of initial and final states, because it is purely virtual, and has a
peak uncertainty, quantified by its width $\Gamma _{\chi }$ divided by two.
The classical limit is Hermitian, like (\ref{LclQEDf}) (see \cite%
{causalityQG}).

The new option is to pursue the strategy of the previous two sections, as in
the extensions (\ref{repl}), (\ref{repl2}) and (\ref{extaQED}). We wish to
interpret $\chi _{\mu \nu }$ as the superposition $\tilde{\chi}_{\mu \nu }+%
\sqrt{2}\chi _{\mu \nu }^{\prime }$ of an extra, observable massive spin-2
particle $\tilde{\chi}_{\mu \nu }$ and a new fakeon $\chi _{\mu \nu
}^{\prime }$.

We recall, from \cite{Absograv}, that the $\chi _{\mu \nu }$ action $S_{\chi
}(g,\phi _{\text{infl}},\chi )$, which can be obtained from (\ref{sqg}) by
means of auxiliary fields, is the sum%
\begin{equation}
S_{\chi }(g,\phi _{\text{infl}},\chi )=-\frac{\lambda }{8\pi G}S_{\text{PF}%
}(g,\chi )+S_{\chi \text{int}}(g,\phi _{\text{infl}},\chi )  \label{scc}
\end{equation}%
of a term proportional to the covariantized Pauli-Fierz action%
\begin{eqnarray}
S_{\text{PF}}(g,\chi ) &=&\frac{1}{2}\int \mathrm{d}^{4}x\sqrt{-g}\left[
D_{\rho }\chi _{\mu \nu }D^{\rho }\chi ^{\mu \nu }-D_{\rho }\chi D^{\rho
}\chi +2D_{\mu }\chi ^{\mu \nu }D_{\nu }\chi -2D_{\mu }\chi ^{\rho \nu
}D_{\rho }\chi _{\nu }^{\mu }\right.  \notag \\
&&\left. -m_{\chi }^{2}(\chi _{\mu \nu }\chi ^{\mu \nu }-\chi ^{2})+R^{\mu
\nu }(\chi \chi _{\mu \nu }-2\chi _{\mu \rho }\chi _{\nu }^{\rho })\right] ,
\label{SPF}
\end{eqnarray}%
with nonminimal terms ($\chi $ denoting the trace $g^{\mu \nu }\chi _{\mu
\nu }$), plus further interactions $S_{\chi \text{int}}(g,\phi _{\text{infl}%
},\chi )$.

The decomposition of $\chi _{\mu \nu }$ in terms of $\tilde{\chi}_{\mu \nu }$
and $\chi _{\mu \nu }^{\prime }$ requires that we treat the quadratic parts
of $\tilde{\chi}_{\mu \nu }$ and $\chi _{\mu \nu }^{\prime }$ differently
from their interactions. We expand the metric tensor $g_{\mu \nu }$ around
the flat-space metric $\eta _{\mu \nu }$ and write 
\begin{equation*}
S_{\chi }(g,\phi _{\text{infl}},\chi )\equiv -\frac{\lambda }{8\pi G}S_{%
\text{PF}}(\eta ,\chi )+\Delta S_{\chi }(\eta ,g,\phi _{\text{infl}},\chi ).
\end{equation*}%
Then we modify the theory according to the strategy of section \ref{PV}. The
interaction part remains the same and contains the combination $\chi _{\mu
\nu }=\tilde{\chi}_{\mu \nu }+\sqrt{2}\chi _{\mu \nu }^{\prime }$. Instead,
the quadratic part is turned into the sum of the quadratic parts of $\chi
_{\mu \nu }^{\prime }$ and $\tilde{\chi}_{\mu \nu }$. Mimicking (\ref{repl}%
), the replacement reads 
\begin{equation}
S_{\chi }(g,\phi _{\text{infl}},\chi )\rightarrow \frac{\lambda }{8\pi G}S_{%
\text{PF}}(\eta ,\tilde{\chi})-\frac{\lambda }{8\pi G}S_{\text{PF}}(\eta
,\chi ^{\prime })+\Delta S_{\chi }(\eta ,g,\phi _{\text{infl}},\tilde{\chi}+%
\sqrt{2}\chi ^{\prime }),  \label{sum}
\end{equation}%
so $\tilde{\chi}_{\mu \nu }$ becomes a physically observable massive spin-2
particle, while $\chi _{\mu \nu }^{\prime }$ must be treated as a fakeon.

The right-hand side of (\ref{sum}) breaks general covariance, because it
depends on both metrics $g_{\mu \nu }$ and $\eta _{\mu \nu }$.\ The fields $%
\tilde{\chi}_{\mu \nu }$ and $\chi _{\mu \nu }^{\prime }$ are defined by
different prescriptions and physically distinguished: $\tilde{\chi}_{\mu \nu
}$, which is a physical particle, must be included in the set of incoming
and outgoing states; $\chi _{\mu \nu }^{\prime }$, as a fakeon, does not
belong there. In these circumstances, it is not obvious how to recover
general covariance. Below we study the issue in more detail.

We remarked in section \ref{PV} that this problem is a well-known aspect of
the Pauli-Villars approach, which treats the interactions differently from
the quadratic parts.

To conclude, the new option, which works well in QED, cannot be used as is
in quantum gravity. This is unfortunate, because the resulting theory would
contain an additional, observable massive spin-2 particle $\tilde{\chi}_{\mu
\nu }$ with respect to the theory of \cite{LWgrav} (as well as a different
spin-2 fakeon $\chi _{\mu \nu }^{\prime }$).

\subsection{General covariance and PV fields}

The breaking of general covariance due to the decomposition of $\chi _{\mu
\nu }$ into the fields $\tilde{\chi}_{\mu \nu }$ and $\chi _{\mu \nu
}^{\prime }$ is entirely due to the quantization prescriptions. For this
reason, the issue deserves a careful analysis.

We begin by describing an alternative procedure to apply the trick of
converting a ghost into the superposition of a physical particle plus a
fakeon, which helps us keep the symmetries under control in a more
transparent way. Let%
\begin{equation*}
S(\phi ,g)=-\frac{1}{2}\int \mathrm{d}^{4}x\hspace{0.01in}[(\partial _{\mu
}\phi )(\partial ^{\mu }\phi )-M^{2}\phi ^{2}]+S_{\text{int}}(\phi ,g)
\end{equation*}%
denote the action of a field $\phi $ with negative kinetic term coupled to
gravity. Adding a decoupled free field $\Omega $ with the same mass, we
obtain 
\begin{equation}
S^{\prime }(\phi ,g,\Omega )=S(\phi ,g)+\frac{1}{2}\int \mathrm{d}^{4}x%
\hspace{0.01in}[(\partial _{\mu }\Omega )(\partial ^{\mu }\Omega
)-M^{2}\Omega ^{2}].  \label{tota}
\end{equation}%
The total action is still invariant under general coordinate
transformations, provided $\Omega $ does not transform. At the infinitesimal
level, the transformations read%
\begin{equation}
\delta \phi =\xi ^{\rho }\partial _{\rho }\phi ,\qquad \delta g_{\mu \nu
}=\xi ^{\rho }\partial _{\rho }g_{\mu \nu }+g_{\mu \rho }\partial _{\nu }\xi
^{\rho }+g_{\nu \rho }\partial _{\mu }\xi ^{\rho },\qquad \delta \Omega =0.
\label{invf}
\end{equation}

The action (\ref{tota}) gives an invariant theory at the quantum level as
long as the quantization prescriptions are compatible with the symmetry (\ref%
{invf}). If so, the Ward-Takahashi-Slavnov-Taylor (WTST) identities \cite%
{STWT} can be derived in the usual fashion, by means of a change of field
variables dictated by (\ref{invf}) in the functional integral, after
introducing the source term $\int \mathrm{d}^{4}x\hspace{0.01in}(J_{\phi
}\phi +J_{\Omega }\Omega +J^{\mu \nu }g_{\mu \nu })$. If we focus on the
matter sector and treat the metric as an external field, the identities read%
\begin{equation}
\int \mathrm{d}^{4}x\hspace{0.01in}\left[ J_{\phi }\langle \delta \phi
\rangle _{J}+J_{\Omega }\langle \delta \Omega \rangle _{J}+J^{\mu \nu
}\delta g_{\mu \nu }\right] =0,  \label{WI}
\end{equation}%
where $\langle \cdots \rangle _{J}$ denotes the connected Green functions at
nonvanishing sources. The term $\langle \delta \Omega \rangle _{J}$ vanishes
by (\ref{invf}). We specialize to $g_{\mu \nu }=\eta _{\mu \nu }$ and $\xi
^{\rho }=$ constant (translations), so that the term $\delta g_{\mu \nu }$
vanishes as well. Differentiating the resulting equation (\ref{WI}) with
respect to $J_{\phi }$ and $J_{\Omega }$ and setting $J_{\phi }=J_{\Omega
}=0 $ afterwards, we find%
\begin{equation}
\hspace{0.01in}\langle \delta \phi (x)\hspace{0.01in}\Omega (y)\rangle =\xi
^{\rho }\hspace{0.01in}\langle \partial _{\rho }\phi (x)\hspace{0.01in}%
\Omega (y)\rangle =0.  \label{clash}
\end{equation}%
In particular, the quantization prescription should not mix $\phi $ with $%
\Omega $: a quadratic contribution like 
\begin{equation}
\int \mathrm{d}^{4}x\mathrm{d}^{4}y\hspace{0.01in}J_{\phi }(x)G_{\text{mix}%
}(x,y)J_{\Omega }(y)  \label{Gmix}
\end{equation}%
to the generating functional of the connected Green functions is not
compatible with general covariance.

Now, observe that the free-field action%
\begin{equation*}
-\frac{1}{2}\int \mathrm{d}^{4}x\hspace{0.01in}[(\partial _{\mu }\phi
)(\partial ^{\mu }\phi )-M^{2}\phi ^{2}]+\frac{1}{2}\int \mathrm{d}^{4}x%
\hspace{0.01in}[(\partial _{\mu }\Omega )(\partial ^{\mu }\Omega
)-M^{2}\Omega ^{2}]
\end{equation*}%
is invariant under the hyperbolic rotation%
\begin{equation}
\phi =\Phi +\sqrt{2}Q,\qquad \Omega =\sqrt{2}\Phi +Q.  \label{trasfa}
\end{equation}%
The rotated action 
\begin{eqnarray}
S^{\prime \prime }(\Phi ,g,Q) &\equiv &S^{\prime }(\Phi +\sqrt{2}Q,g,\sqrt{2}%
\Phi +Q)=\frac{1}{2}\int \mathrm{d}^{4}x\hspace{0.01in}[(\partial _{\mu
}\Phi )(\partial ^{\mu }\Phi )-M^{2}\Phi ^{2}]  \notag \\
&&-\frac{1}{2}\int \mathrm{d}^{4}x\hspace{0.01in}[(\partial _{\mu
}Q)(\partial ^{\mu }Q)-M^{2}Q^{2}]+S_{\text{int}}(\Phi +\sqrt{2}Q,g)
\label{rota}
\end{eqnarray}%
matches the actions (\ref{repl}), (\ref{repl2}), (\ref{extaQED}) and (\ref%
{sum}). It is precisely what we need to decompose the field $\phi$ into a
physical particle $\Phi $ plus a fakeon $Q$.

We can read the symmetries of $S^{\prime \prime }(\Phi ,g,Q)$ from (\ref%
{invf}). They are%
\begin{equation*}
\delta \Phi =-\xi ^{\rho }\partial _{\rho }\Phi -\sqrt{2}\xi ^{\rho
}\partial _{\rho }Q,\qquad \delta Q=-\sqrt{2}\delta \Phi .
\end{equation*}%
The free-field propagators we want can be derived from (\ref{pp2}). They
are, in momentum space,%
\begin{equation}
\langle \Phi (p)\hspace{0.01in}\Phi (-p)\rangle _{0}=\frac{i}{%
p^{2}-M^{2}+i\epsilon },\quad \langle \Phi (p)\hspace{0.01in}Q(-p)\rangle
_{0}=0,\quad \langle Q(p)\hspace{0.01in}Q(-p)\rangle _{0}=-\left. \frac{i}{%
p^{2}-M^{2}}\right\vert _{\text{f}}.  \label{ffp}
\end{equation}%
Switching to the field variables $\phi $ and $\Omega $ by means of (\ref%
{trasfa}), we find, for legs that disconnect the diagrams,%
\begin{eqnarray}
\langle \phi (p)\hspace{0.01in}\phi (-p)\rangle _{0} &=&-\frac{i}{%
p^{2}-M^{2}-i\epsilon },\qquad \langle \phi (p)\hspace{0.01in}\Omega
(-p)\rangle _{0}=\sqrt{2}\pi \delta (p^{2}-M^{2}),\qquad  \notag \\
\langle \Omega (p)\hspace{0.01in}\Omega (-p)\rangle _{0} &=&\frac{i}{%
p^{2}-M^{2}+i\epsilon }+\pi \delta (p^{2}-M^{2}).  \label{2pt}
\end{eqnarray}

We see that, although $\langle \phi (p)\hspace{0.01in}\phi (-p)\rangle _{0}$
is the desired one, i.e. (\ref{-iep}), we cannot fulfil (\ref{clash}) and
make $G_{\text{mix}}$ vanish. This is inconsistent with the WTST identities.
Moreover, the $Q$ projection amounts to set $J_{Q}=0$, which does not kill
the contributions like (\ref{Gmix}). For these reasons, we cannot ensure
that general covariance can be recovered.

The case of gravity is obtained by means of the substitutions $\phi
\rightarrow \chi _{\mu \nu }$, $\Phi \rightarrow \tilde{\chi}_{\mu \nu }$, $%
Q\rightarrow \chi _{\mu \nu }^{\prime }$, and adapting the formulas where
necessary. The role of $\Omega $ is played by a free Pauli-Fierz spin-2
particle $\Omega _{\mu \nu }$ of mass $m_{\chi }$. Renormalizability is
ensured by the very fact that $\Omega _{\mu \nu }$ decouples from the rest
(apart from the quantization prescription, which does not affect the
renormalizability). The conclusions do not change.

\subsection{Manifestly covariant decomposition by means of a massive spin-2
multiplet}

A way to perform the decomposition in a manifestly covariant way is to
include a Pauli-Fierz spin-2 particle $\Omega _{\mu \nu }$ of mass $m_{\chi
} $, coupled to gravity as required by general covariance, and then rotate
the degenerate pair $\chi _{\mu \nu }$, $\Omega _{\mu \nu }$, so as to
single out the physically observable spin-2 particle $\tilde{\chi}_{\mu \nu
} $ and the fakeon $\chi _{\mu \nu }^{\prime }$. However, such a theory is
not renormalizable, because the Pauli-Fierz propagator does not fall off as
required by power counting at large momenta.

It is possible to have renormalizability (and unitarity) if we replace $%
\Omega _{\mu \nu }$ with a whole massive spin-2 multiplet $\Upsilon _{\mu
\nu }$, of the type studied in ref. \cite{HS}. In that case $\Upsilon _{\mu
\nu }$ is a symmetric, traceless tensor and contains a triplet: the spin-2
particle $\Omega _{\mu \nu }$, a spin-1 fakeon $\Omega _{\mu }$ and a
massive scalar $\Omega $. If we choose the mass of $\Omega _{\mu \nu }$ to
be equal to $m_{\chi }$, to have degeneracy with $\chi _{\mu \nu }$, the
masses $m_{1}$ and $m_{0}$ of $\Omega _{\mu }$ and $\Omega $ are related to $%
m_{\chi }$ by a certain formula that can be found in \cite{HS}. Then, we
assume that $\chi _{\mu \nu }$ has the free propagator of a LW\ ghost and
complete the set of free propagators as in (\ref{2pt}), so that, after
rotating the degenerate pair $\chi _{\mu \nu }$, $\Omega _{\mu \nu }$, we
can identify the physically observable spin-2 particle $\tilde{\chi}_{\mu
\nu }$ and the fakeon $\chi _{\mu \nu }^{\prime }$, with free propagators of
the form (\ref{ffp}).

So doing, we manage to extend the original LW concept to gravity in a
general covariant way, under the requirement that the projected classical
action be Hermitian. However, what we obtain is just the theory of \cite%
{LWgrav} coupled to matter in a peculiar way.

\section{Conclusions}

\label{conclusions}\setcounter{equation}{0}

The Lee-Wick models rely on the premise that a unitary reduced $S$ matrix $%
S_{r}$ can be built by removing the LW ghosts, which are unstable,\ from the
sets of asymptotic states. However, a finite lifetime is not a sufficient
reason to ignore a particle from the physical spectrum. If we just drop the
LW ghosts, saving the muon and the resonances, the models have non-Hermitian
classical limits.

A proper classical limit is important to develop a meaningful cosmology.
Although it is legitimate to ignore heavy massive particles at low energies
in particle physics, in primordial cosmology the understanding of high
energies (subhorizon scales) is necessary to make predictions about the low
energies (superhorizon scales). The Bunch-Davies condition \cite{BD,baumann}%
, for example, specifies the vacuum in the subhorizon region. We cannot make
realistic assumptions about that region, which is experimentally and
observationally inaccessible, if the theory has ghosts or non-Hermitian
interactions. The ABP bound $m_{\chi }>m_{\phi }/4$ of \cite{ABP}, crucial
for the prediction of the tensor-to-scalar ratio $r$, also follows from the
interpolation between the subhorizon and the superhorizon scales.

We have shown that a nonpurely virtual particle cannot be completely
removed, within the realm of perturbation theory. Barring nonperturbative
mechanisms, unacceptable remnants emerge one way or another, like a
non-Hermitian self-interaction, an indefinite metric or a Hamiltonian that
is unbounded from below.

Fakeons, on the other hand, are purely virtual, so it is not necessary to
worry about making them decay. For this reason, they avoid the problems of
the other options, without using semiperturbative approaches or advocating
nonperturbative effects. Besides being fully perturbative, the models with
fakeons have a Hermitian classical limit and a Hermitian reduced action. In
quantum gravity, they lead to a predictive primordial cosmology. The fakeon
width $\Gamma $ is not interpreted as a lifetime, but as (twice) the
magnitude of the peak\ uncertainty, for processes that probe energies close
to the fakeon mass.

The investigation carried out in this paper suggests a way to remove a LW
ghost only partially, after converting it into a superposition of a fakeon
and an observable physical particle. Under certain assumptions, this trick
makes the Pauli-Villars fields consistent without sending their masses to
infinity. It also allows us to build a finite QED. Nevertheless, it works
only with neutral matter fields, in the absence of gravity, because it
clashes with general covariance and gauge invariance. A manifestly covariant
decomposition can be obtained by adding a massive spin-2 multiplet, which in
the end just gives quantum gravity coupled to matter in a peculiar way.

\vskip12truept \noindent {\large \textbf{Acknowledgments}}

\vskip 2truept

We are grateful to D. Comelli, E. Gabrielli and M. Piva for helpful
discussions. This work was supported in part by the European Regional
Development Fund through the CoE program grant TK133 and the Estonian
Research Council grant PRG803.

\end{document}